\documentclass{pasj00}

\usepackage{times}
\begin{document}

\SetRunningHead{M. Choi et al.}{NGC 1333 IRAS 4A Imaging}

\title{Radio Imaging of the NGC 1333 IRAS 4A Region:
       Envelope, Disks, and Outflows of a Protostellar Binary System}
\author{Minho \textsc{Choi}\altaffilmark{1,2},
        Miju \textsc{Kang}\altaffilmark{1},
        Ken'ichi \textsc{Tatematsu}\altaffilmark{3},
        Jeong-Eun \textsc{Lee}\altaffilmark{4},
        and
        Geumsook \textsc{Park}\altaffilmark{5}}
\altaffiltext{1}{Korea Astronomy and Space Science Institute,
                 776 Daedeokdaero, Yuseong, Daejeon 305-348, South Korea}
\email{minho@kasi.re.kr}
\altaffiltext{2}{University of Science and Technology,
                 217 Gajeongro, Yuseong, Daejeon 305-350, South Korea}
\altaffiltext{3}{National Astronomical Observatory of Japan,
                 2-21-1 Osawa, Mitaka, Tokyo 181-8588, Japan}
\altaffiltext{4}{Department of Astronomy and Space Science,
                 Kyung Hee University, Yongin, Gyeonggi 446-701, South Korea}
\altaffiltext{5}{Department of Physics and Astronomy,
                 Seoul National University, Seoul 151-742, South Korea}


\maketitle

\begin{abstract}
The NGC 1333 IRAS 4A protobinary was observed
in the 1.3 cm and 6.9 mm continuum and the ammonia and SiO lines,
with an angular resolution of about 0.4 arcseconds.
The continuum maps show the circumstellar structures
of the two protostars, A1 and A2.
The A1 system is brighter and more massive than the A2 system.
The ratio of mass, including dense gas and protostar, is about 6.
The properties of the circumstellar disks and outflows suggest
that A1 may be younger than A2.
The deflected part of the northeastern jet of A2
is bright in the SiO line,
and the distance between the brightest peak and deflection point
suggests that the enhancement of SiO takes about 100 yr
after the collision with a dense core.
The ammonia maps show a small structure
that seems to be a part of the obstructing core.
The outflow properties were studied by comparing interferometric maps
of SiO, ammonia, formaldehyde, and HCN lines.
Their overall structures agree well,
suggesting that these species are excited by the same mechanism.
However, the intensity distributions show that SiO is chemically unique.
SiO may be directly linked to the primary jet
while the other species may be tracing the entrained ambient gas.
\end{abstract}

\setcounter{footnote}{5}

\section{INTRODUCTION}

NGC 1333 IRAS 4A2 is one of the best-studied low-mass protostars.
It has a circumstellar disk viewed nearly edge-on
and drives a well-collimated and rotating outflow/jet
(Choi 2005; Choi et al. 2007, 2011).
Choi et al. (2010) found
that the NH$_3$ (2, 2) and (3, 3) lines
seem to selectively trace the accretion disk
while the millimeter/centimeter continuum
seems to trace both disk and envelope.
The rotation kinematics of the disk suggests
that the mass of the central protostar is $\sim$0.08 $M_\odot$
and that the collapse age is $\sim$50,000 yr (Choi et al. 2010).
The bipolar jet flows in the northeast-southwest direction
(Blake et al. 1995; Hodapp \& Ladd 1995; Lefloch et al. 1998).
The jet may be launched from a small region on the disk,
or outflow foot-ring, with a radius of $\sim$2 AU (Choi et al. 2011).
The northeastern jet shows a sharp bend of flow direction
that may be caused by a collision with a dense core in the ambient cloud
(Choi 2005; Baek et al. 2009).
IRAS 4A2 also exhibits other star formation activities
such as H$_2$O maser emission (Park \& Choi 2007; Marvel et al. 2008).

IRAS 4A2 belongs to a binary system (IRAS 4A)
in the Perseus star-forming region at a distance of 235 pc from the Sun
(Lay et al. 1995; Looney et al. 2000; Hirota et al. 2008).
The spectral energy distribution of IRAS 4A suggests
that the constituent binary members are Class 0 protostars
(Sandell et al. 1991; Enoch et al. 2009).
The luminosity of IRAS 4A suggests
that the protostars are growing at a rate typical of Sun-like stars
through the accretion of gas from the molecular envelope
(Jennings et al. 1987; Choi et al. 2010).

IRAS 4A1 is brighter in the millimeter/centimeter continuum
than its sibling protostar A2 (Looney et al. 2000;
Reipurth et al. 2002; J{\o}rgensen et al. 2007),
but it is weaker in the NH$_3$ emission (Choi et al. 2007).
IRAS 4A1 seems to drive an outflow to the south,
but its counter flow, if any, has not been detected clearly (Choi 2005).
While A1 must be in an evolutionary stage similar to that of A2,
the details are less clear.

In this paper, we present the results of
our observations of the NGC 1333 IRAS 4A region
in the 1.3 cm and 6.9 mm continuum, the NH$_3$ lines, and the SiO line
with an angular resolution higher than those of the previous studies
(Choi 2005; Choi et al. 2007).
We describe our observations in Section 2.
In Section 3, we report the results
and discuss the star-forming activities in the IRAS 4A region.
A summary is given in Section 4.

\section{OBSERVATIONS AND DATA}

\subsection{K-band Observations}

\begin{figure*}[!t]
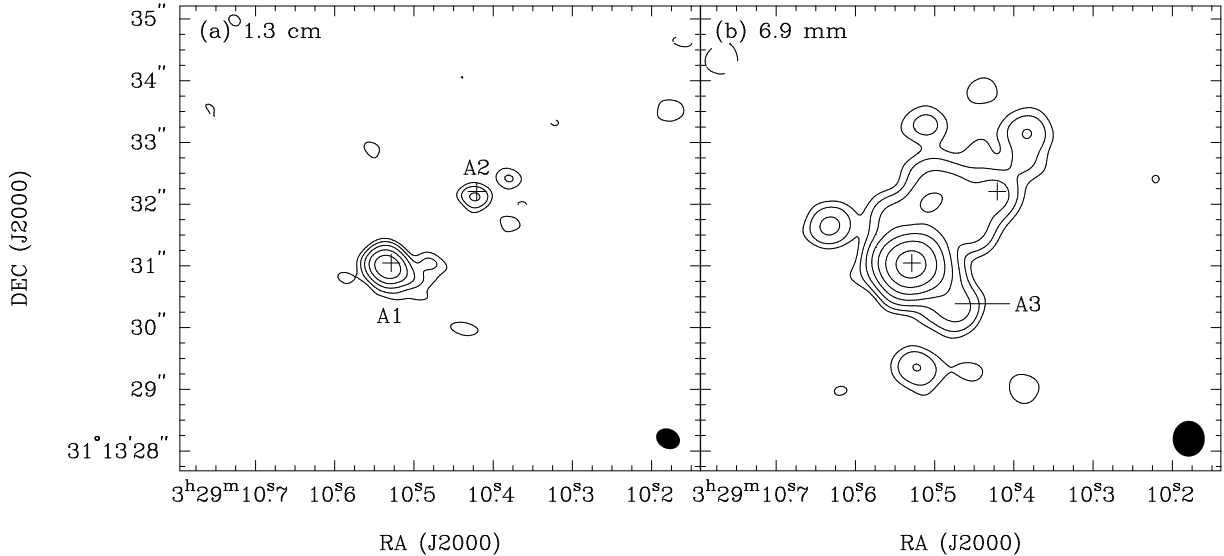

\begin{center}
\FigureFile(162mm,0){f01.eps}
\end{center}
\caption{
Maps of the continuum emission toward the NGC 1333 IRAS 4A region.
The contour levels are 1, 2, 4, 8, 16, 32, and 64 $\times$ 0.05 mJy beam$^{-1}$.
Dashed contours are for negative levels.
(a)
Map of the $\lambda$ = 1.3 cm continuum made using the B-array data only.
The rms noise is 0.016 mJy beam$^{-1}$.
Shown in the bottom right-hand corner is the synthesized beam:
FWHM = 0\farcs40 $\times$ 0\farcs33 with PA = 64$^\circ$.
(b)
Map of the $\lambda$ = 6.9 mm continuum.
A relatively bright substructure is labeled as A3.
The rms noise is 0.017 mJy beam$^{-1}$.
The synthesized beam has
FWHM = 0\farcs58 $\times$ 0\farcs52 with PA = --1$^\circ$.
Plus signs:
the 3.6 cm continuum sources (Reipurth et al. 2002).}
\end{figure*}

The NGC 1333 IRAS 4 region was observed using the Very Large Array (VLA)
of the National Radio Astronomy Observatory
in the NH$_3$ (2, 2) and (3, 3) lines
and in the $\lambda$ = 1.3 cm continuum.
Observations in the C-array configuration
were presented in Choi et al. (2007).
Details of the observations in the B-array configuration
and results for the NH$_3$ lines
were given in Choi et al. (2010).
The continuum was observed for 15 minutes at the beginning
and for another 15 minutes at the end of each observing track,
and the NH$_3$ lines were observed in the midsection of the track.

For the 1.3 cm continuum, 
the observations were made
in the standard K-band continuum mode (22.5 GHz).
Maps were made using a CLEAN algorithm.
With a natural weighting,
the B-array continuum data produced a synthesized beam
of 0\farcs40 $\times$ 0\farcs33 in full-width at half-maximum (FWHM).

\subsection{Q-band Observations}

The NGC 1333 IRAS 4 region was observed using the VLA
in the SiO $v=0$ $J$ = 1 $\rightarrow$ 0 line (43.423858 GHz)
and in the $\lambda$ = 6.9 mm continuum.
Twenty-five antennas were used in the C-array configuration on 2004 March 1.
The continuum was observed for 20 minutes at the beginning
and for 10 minutes at the end of each observing track,
and the SiO line was observed in the midsection of the track.
For the SiO line, the spectral windows were set to have 64 channels
with a channel width of 0.098 MHz,
giving a velocity resolution of 0.67 km s$^{-1}$.
For the 6.9 mm continuum, 
the observations were made
in the standard Q-band continuum mode (43.3 GHz).

The phase tracking center was ($\alpha$, $\delta$)
= (\timeform{03h29m10.993s}, \timeform{31D13'45.21''}) in J2000.0.
Note that the field of view of the C-array observations corresponds
to the first field of the D-array mosaic observations (Choi 2005),
which covers the northeastern part of the IRAS 4A area
including the redshifted outflow of IRAS 4A2.
The southwestern field was not covered in the C-array observations,
and most part of the blueshifted outflows are outside the field of view.

The nearby quasar 0336+323 (PKS 0333+321) was observed
to determine the phase and to obtain the bandpass response.
The flux was calibrated by observing the quasar 0713+438 (QSO B0710+439)
and by setting its flux density to 0.20 Jy
(VLA Calibrator Flux Density Database%
\footnote{See http://aips2.nrao.edu/vla/calflux.html.}).
To check the flux scale,
the quasar 0319+415 (3C 84) was observed in the same track.
Comparison of the amplitude gave a flux density of 4.28 Jy for 0319+415,
which agrees with the value in the VLA Calibrator Flux Density Database.
The bootstrapped flux density of 0336+323 was 1.26 Jy.
To avoid the degradation of sensitivity owing to pointing errors,
pointing was referenced
by observing the calibrators at the X band ($\lambda$ = 3.6 cm).
This referenced pointing was performed
about once an hour and just before observing the flux calibrator.
With a natural weighting,
the SiO data produced a synthesized beam
of FWHM = 0\farcs55 $\times$ 0\farcs52,
and the continuum data produced a beam
of FWHM = 0\farcs58 $\times$ 0\farcs52.

\section{RESULTS AND DISCUSSION}

\subsection{Continuum Emission Maps}

\begin{table*}
\caption{NGC 1333 IRAS 4A Continuum Source Parameters}
\begin{center}
\begin{tabular}{lcccccccc}
\hline
& \multicolumn{2}{c}{Peak Position$^a$}
&& \multicolumn{2}{c}{1.3 cm Flux Density$^b$}
&& \multicolumn{2}{c}{6.9 mm Flux Density$^b$} \\
\cline{2-3} \cline{5-6} \cline{8-9}
Source & $\alpha_{\rm J2000.0}$ & $\delta_{\rm J2000.0}$
&& Peak & Total && Peak & Total  \\
\hline
A1 & 03 29 10.53 & 31 13 31.0 &
   & 1.554 $\pm$ 0.016 & 2.47 $\pm$ 0.07 &
   & 5.57  $\pm$ 0.02  & \phantom{0}9.51 $\pm$ 0.12 \\
A2 & 03 29 10.42 & 31 13 32.1 &
   & 0.225 $\pm$ 0.016 & 0.37 $\pm$ 0.07 &
   & 0.39  $\pm$ 0.02  & \phantom{0}1.46 $\pm$ 0.08 \\
\hline
A  & \ldots & \ldots &
   & \ldots & 2.66 $\pm$ 0.18 &
   & \ldots  & 11.5\phantom{0} $\pm$ 0.3\phantom{0} \\
\hline
\end{tabular}
\parbox{150mm}{\medskip
$^a$ Source positions are from the 1.3 cm continuum map.
     Units of right ascension are hours, minutes, and seconds,
     and units of declination are degrees, arcminutes, and arcseconds. \\
$^b$ Flux densities are in mJy beam$^{-1}$ or in mJy,
     corrected for the primary beam response.
     The box sizes used for measuring the total flux
     are the same as those listed in Table 2 of Looney et al. (2000):
     2\farcs9 $\times$ 2\farcs2 for A1, 1\farcs8 $\times$ 1\farcs6 for A2,
     and 5\farcs4 $\times$ 6\farcs2 for A.}
\end{center}
\end{table*}

Figure 1 shows the continuum emission maps,
and source parameters are listed in Table 1.
The 1.3 cm continuum map shows the compact structures of IRAS 4A1 and A2
as well as a few weak features.
IRAS 4A1 is marginally resolved,
and an elliptical Gaussian fit gives
a deconvolved size of FWHM = 0\farcs24 $\times$ 0\farcs16
with a position angle (PA) of 52$^\circ$.
IRAS 4A2 is unresolved.
The 6.9 mm continuum map shows IRAS 4A1, A2,
and an extended structure surrounding the binary system.
IRAS 4A2 does not stand out
because the extended structure has a comparable brightness.
Some of the emission peaks in the extended structure
are located away from A1 or A2
with angular distances comparable to the A1-A2 binary separation,
which suggests that there is a common envelope around the binary system.

The measurement of total flux is a nontrivial issue,
especially for the 6.9 mm map,
because the source structure is complicated.
The total flux of A1 is not very sensitive
to the choice of integration area (a box in this case)
because the main peak of A1 is much brighter
than the other features in the area.
The total flux of A2 at 6.9 mm, however, is sensitive to the choice of box
because of the extended emission around A2.
Since comparisons with the 2.7 mm flux densities are important (see below),
we measured the total flux densities (listed in Table 1)
in the boxes used by Looney et al. (2000),
for consistency in measuring spectral index.
However, the A1 and A2 boxes have an overlapping region,
and the A box is larger than the union of A1 and A2 boxes.
Therefore, the flux density of A is not a simple sum of those of A1 and A2.
In the calculation of mass given below,
we mostly rely on the 1.3 cm and 2.7 mm total flux densities
of the whole IRAS 4A region,
and the ambiguity in the 6.9 mm flux density of A2 is not critical.

\begin{figure}[!b]
\begin{center}
\FigureFile(79mm,0){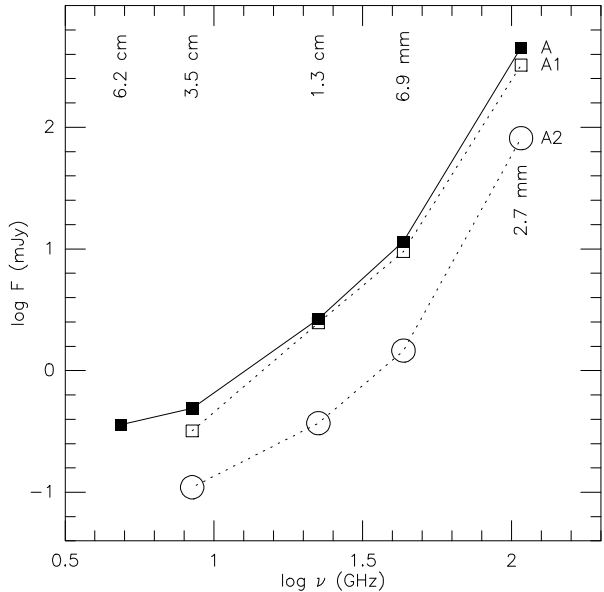}
\end{center}
\caption{
Spectral energy distributions.
Filled squares: IRAS 4A (including A1, A2, and common envelope).
Open squares: IRAS 4A1.
Open circles: IRAS 4A2.
Flux densities are from Rodr{\'\i}guez et al. (1999), Reipurth et al. (2002),
this work, and Looney et al. (2000).
The flux uncertainties are smaller than the size of markers.}
\end{figure}

Figure 2 shows the spectral energy distributions (SEDs)
in the centimeter-millimeter wavelength region.
SEDs can be described using a power-law form,
$F \propto \nu^\alpha$, where $F$ is the flux density,
$\nu$ is the frequency, and $\alpha$ is the spectral index.
Dust emission from a disk/envelope system
usually has a large spectral index ($\alpha \gtrsim 2$),
and free-free emission from a thermal radio jet
usually has a small spectral index ($\alpha \lesssim 1$)
(Reynolds 1986; Anglada et al. 1998).
The SED of the IRAS 4A system shows
that the slope is steep at shorter wavelengths
and shallow at longer wavelengths.
The spectral index is large ($\alpha \gtrsim 1.9$)
at wavelengths shorter than 3.5 cm,
and small ($\alpha \approx 0.6$) in the 6.2--3.5 cm section.
This change of $\alpha$ around 3.5 cm suggests
that the continuum emission at 1.3 cm and shorter wavelengths
mostly comes from dust
and that the 6.2 cm continuum may come from free electrons.
Even at 3.5 cm, a significant fraction of the flux may come from dust.

The mass of molecular gas can be estimated from the SED of the dust continuum
using the mass emissivity given by Beckwith \& Sargent (1991),
\begin{equation}
   \kappa_\nu = 0.1 \left({\nu\over{\nu_0}}\right)^\beta
                {\rm cm^2\ g^{-1}},
\end{equation}
where $\nu_0$ = 1200 GHz, and $\beta$ is the opacity index.
For the 1.3 cm -- 2.7 mm section,
the opacity index of IRAS 4A is $\beta \approx 1.3$.
Assuming optically thin emission, the mass can be estimated by
\begin{equation}
   M = {{F_\nu D^2}\over{\kappa_\nu B_\nu(T_d)}},
\end{equation}
where $D$ is the distance to the source, $B_\nu$ is the Planck function,
and $T_d$ is the dust temperature.

Assuming $D$ = 235 pc and $T_d$ = 33 K
(Hirota et al. 2008; Jennings et al. 1987),
the mass of IRAS 4A is $M$ = 2.6 $\pm$ 0.8 $M_\odot$.
The mass estimate is sensitive to the value of $\beta$.
For example, Looney et al. (2000) derived a mass of 1.3 $M_\odot$
(scaled to the distance of 235 pc)
assuming $\beta$ = 1 and using the 2.7 mm flux density only.
Therefore, the uncertainty in mass introduced by the assumptions on $\beta$
may be about a factor of 2.
This mass is based on the flux densities
in a box of 5\farcs4 $\times$ 6\farcs2 around IRAS 4A1,
which may include the circumstellar disk and protostellar envelope around A1,
those around A2, and the common envelope.
For comparison, the mass of the whole protostellar envelope of IRAS 4A
derived from single-dish observations
is $\sim$7 $M_\odot$ (Enoch et al. 2009, scaled to 235 pc),
and the mass of the IRAS 4A2 protostar
derived from the kinematics of the circumstellar disk
is $\sim$0.08 $M_\odot$ (Choi et al. 2010).

The mass of molecular gas around each protostar
can be derived using the ratio of flux densities,
assuming that the dust properties are nearly uniform
in the whole IRAS 4A system.
IRAS 4A1 and A2 are most clearly separated in the 1.3 cm continuum map,
and the 1.3 cm flux densities give
2.3 $\pm$ 0.7 $M_\odot$ for A1 and 0.3 $\pm$ 0.1 $M_\odot$ for A2.
Then the total mass of the A2 system (protostar and dense molecular gas)
is $\sim$0.4 $M_\odot$.
Therefore, regardless of the mass of the A1 protostar,
the A1 system is much more massive than the A2 system.

The extended structure in the 6.9 mm continuum map (Figure 1b)
is clumpy and shows several substructures.
The overall morphology of the extended structure
is similar to what was seen in the 2.7 mm continuum
(see Figure 12d of Looney et al. 2000).
One of the substructures is relatively bright
and can be seen in both the 6.9 mm and 2.7 mm maps.
This substructure, labeled as A3 in Figure 1b,
has a peak position at 0\farcs7 west and 0\farcs7 south with respect to A1.
The nature of the substructures is not clear.
Some of them may be dense clumps in the envelope
(that may eventually fall into one of the protostars),
some may be prestellar cores,
and some might even harbor small compact objects.

\subsection{NH$_3$ Line Maps of the IRAS 4A1 Region}

\begin{figure}[!b]
\begin{center}
\FigureFile(79mm,0){f03.eps}
\end{center}
\caption{
Maps of the NH$_3$ lines toward IRAS 4A1 from the VLA B-array data.
Contribution from the continuum emission was subtracted out.
(a)
Map of the (2, 2) line core, $V_{\rm LSR}$ = (5.8, 7.6) km s$^{-1}$.
(b)
Maps of the blueshifted, (4.5, 5.8) km s$^{-1}$,
and redshifted, (7.6, 8.9) km s$^{-1}$, parts of the (2, 2) line.
(c)
Map of the (3, 3) line core, (5.8, 7.6) km s$^{-1}$.
(d)
Maps of the blueshifted, (4.6, 5.8) km s$^{-1}$,
and redshifted, (7.6, 8.8) km s$^{-1}$, parts of the (3, 3) line.
Contour levels are 3, 4, and 5 times the rms noise
[0.35 mJy beam$^{-1}$ for ($a$) and ($c$),
and 0.43 mJy beam$^{-1}$ for ($b$) and ($d$)].
Shown in the bottom right-hand corner is the restoring beam: FWHM = 0\farcs3.
Cross:
peak position of the 1.3 cm continuum.
Arrow:
direction of the southern outflow (toward outflow peak 16).}
\end{figure}

\begin{figure}[!t]
\begin{center}
\FigureFile(79mm,0){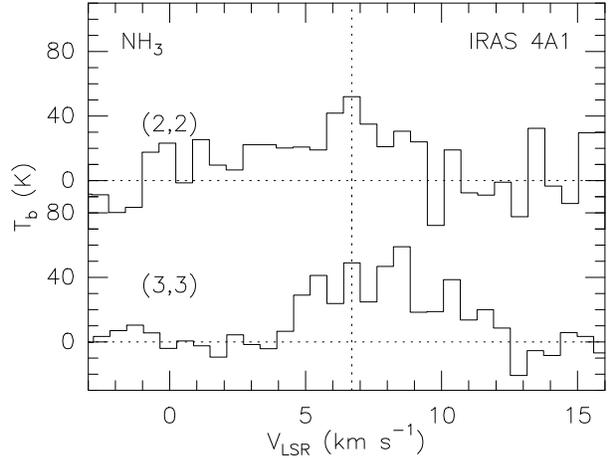}
\end{center}
\caption{
Spectra of the NH$_3$ lines toward IRAS 4A1,
at the peak position of the 1.3 cm continuum emission.
Vertical dotted line:
systemic velocity of IRAS 4A ($V_{\rm LSR}$ = 6.7 km s$^{-1}$; Choi 2001).}
\end{figure}

Figure 3 shows high-resolution (B-array) NH$_3$ maps
of the region around IRAS 4A1,
and Figure 4 shows the spectra.
(See Choi et al. 2007 for NH$_3$ maps in a lower resolution.)
The source structure and spectra of A1 is somewhat different from those of A2
(see Choi et al. 2010 for high-resolution maps of the A2 region).
In the case of A2, the NH$_3$ lines seem to trace the circumstellar disk
since the source shape is elongated
in the direction perpendicular to the outflow
and the spectra has a profile consistent with a rotating disk
(Choi et al. 2010).
By contrast, the NH$_3$ maps of A1 show a more complicated morphology,
and the spectra of A1 show relatively weak line cores and wider line wings
than those of A2.
The total integrated NH$_3$ flux of A1 is smaller than that of A2
by a factor of 1.7.

The (2, 2) line core map (Figure 3a) shows
a structure associated with the continuum source.
This structure (central component) is elongated in the east-west direction,
which may be the circumstellar disk,
considering that the blueshifted outflow of A1 is directed toward the south
(Choi 2005).
The (3, 3) line maps (Figures 3c and 3d)
seem to show the central component
and a secondary structure about 0\farcs4 south (southern component).
The southern component is blueshifted
and may be related with the southern outflow.
The peak positions of the blueshifted and redshifted line wings
of the central component (Figure 3d)
could not be separated with the 0\farcs3 beam,
and the rotation of the A1 disk, if any,
may be slower than that of the A2 disk.
Therefore, the rotation curve could not be analyzed with our data.

In summary, the circumstellar structure of A1 traced by the NH$_3$ lines
seems to be a mixture of the accretion disk and the southern outflow.
The A1 disk is weaker than the A2 disk in the NH$_3$ lines.
The rotation of A1 disk seems to be slower than that of A2,
suggesting that the A1 protostar is less massive than the A2 protostar.

\subsection{Evolutionary Status of the IRAS 4A System}

The IRAS 4A system is one of the youngest and best-studied
protostar binary systems.
The evolutionary status of A2 is especially well-studied
thanks to the rotation kinematics of the circumstellar disk
(Choi et al. 2010).
An interesting issue is the evolutionary status of A1
in comparison with A2.

Previously, Choi et al. (2007) examined the IRAS 4A system
with NH$_3$ and continuum images of $\sim$1$''$ resolution.
To understand the difference in the NH$_3$-to-continuum flux ratios
they explored two possibilities:
(1) the protostars are exactly coeval,
or (2) A2 is a protostar,
and A1 may be a transitionary object on the verge of protostellar collapse.
Because the second possibility contradicts other observational facts,
they preferred the first possibility.
However, it implied that the dust-to-gas ratios of A1 and A2
may be quite different, by almost an order of magnitude,
which is not easy to understand.

The new higher-resolution images presented in this paper suggest
that the difference in the properties of the disks
may be the key in understanding the NH$_3$-to-continuum flux ratios
and also the evolutionary status of the IRAS 4A system.
Here we revisit the issue and explore the possibility
that A1 and A2 are protostars at slightly different stages of evolution.
That is, though A1 and A2 are members of a single binary system,
their collapse ages can be different, i.e., not exactly coeval.
Then which one is the younger of the two?
We suggest that A1 is younger than A2 for several reasons.

First, the accretion disk of A2 traced in the NH$_3$ lines
shows a clearly detectable rotation kinematics,
while the A1 disk does not.
Since the disk around more massive object rotates faster,
the A2 protostar may be more massive than the A1 protostar.
Therefore, if the accretion rates are similar,
the A1 protostar may be younger than the A2 protostar.
(If the accretion rate of A1 is higher,
as the envelope of A1 is more massive,
then the age of A1 may be even much smaller than that of A2.)

Second, the A2 disk is brighter than the A1 disk in the NH$_3$ lines.
The strength of the NH$_3$ lines
is likely correlated with the mass of the disk,
and the disk mass increases monotonically
in the simplest models of protostellar evolution (Young \& Evans 2005).
Therefore, the difference in the NH$_3$ line strength implies
that, if the accretion rates are similar,
the A1 disk is less massive and younger than the A2 disk.

Third, the outflow of A2 is much longer than that of A1.
The length of the northeast-southwest bipolar outflow driven by A2
is $\sim$130$''$ (from the A2 protostar to the H$_2$ knot HL3;
see Figure 1 of Choi et al. 2006).
The length of the southern outflow driven by A1 is $\sim$15$''$
(from the A1 protostar to the SiO outflow peak 16;
see Figure 2a of Choi 2005).
Assuming a proper motion of 0\farcs064 yr$^{-1}$ (Choi et al. 2006)
for rough estimates,
the time scale of the A1 outflow
is shorter than that of A2 by $\sim$1,800 yr.

The relative youth of A1,
despite its relatively massive circumstellar material,
suggests that the progenitor prestellar cores evolved at different speeds.
As a simplest scenario of the evolution of binary prestellar cores,
suppose that the parent cloud fragmented into two cores:
core 1 with a mass of $M_{C1}$ that later evolves to become the A1 system,
and core 2 with a mass of $M_{C2}$ to become the A2 system.
If the collapse age of A2 is indeed larger than that of A1,
core 2 may have evolved faster and formed a protostar earlier.
Since the denser core evolves faster
(that is, the evolution time scale is shorter for the denser core),
core 2 may have had a higher density than core 1.
This scenario can be understood with a simple argument.
At the time of the fragmentation,
suppose that each of the two cores has a roughly spherical shape
(or any shape similar to each other),
and the boundary (the point of contact between the two spheres) is
the point where the gravitational forces of the two cores balance out.
Then the ratio of radii is $R_{C1}/R_{C2} \approx \sqrt{M_{C1}/M_{C2}}$,
and the density ratio is $n_{C1}/n_{C2} \approx 1 / \sqrt{M_{C1}/M_{C2}}$.
Though the exact density ratio
may depend on details of the fragmentation process,
the more massive core may usually have a smaller average density
than the sibling core.
(For the IRAS 4A system, if the mass ratio at the time of fragmentation
is similar to the current value,
$M_{C1} / M_{C2} \approx M_{A1} / M_{A2} \approx 6$
and $n_{C1}/n_{C2} \approx 0.4$.)
Therefore, it is reasonable that, when a cloud fragments into two cores,
the less massive one would be denser, evolve faster,
and form a protostar earlier.
Considering the amount of circumstellar material, however,
A1 may outgrow A2 in the future
and eventually become a more massive star.

\begin{figure}[!b]
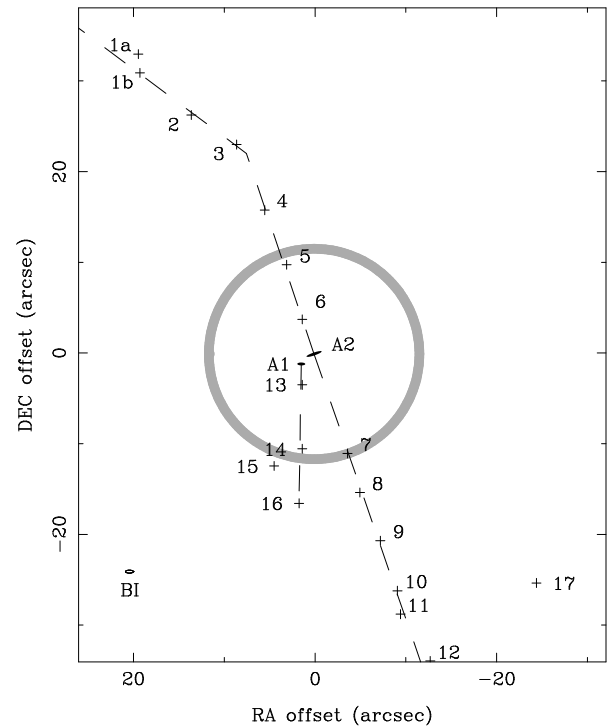

\begin{center}
\FigureFile(79mm,0){f05.eps}
\end{center}
\caption{
Schematic diagram of the IRAS 4A system.
Protostars A1, A2, and BI are labeled.
Plus signs:
SiO outflow peaks.
See Figure 2a of Choi (2005).
Dashed lines:
outflow axes.
Gray circle:
infall region with a radius of 2,700 AU,
which corresponds to the size of the spherical collapse wave
with a sound speed of 0.27 km s$^{-1}$ and a collapse age of 50,000 yr.}
\end{figure}

\begin{figure*}[!t]
\begin{center}
\FigureFile(162mm,0){f06.nogray.eps}
\end{center}
\centerline{\scriptsize
[See http://minho.kasi.re.kr/Publications.html for the original figure.]}
\vspace{-\baselineskip}
\caption{
Maps of the NH$_3$ (3, 3) line toward the IRAS 4A region,
made from the VLA C-array data (also see Figure 1 of Choi et al. 2007).
The maps were convolved to have an angular resolution of FWHM = 2\farcs0
as shown in the bottom left-hand corner.
For these maps, the continuum levels were not subtracted from the line data,
and the structure associated with IRAS 4A1, A2, and BI
comes from both the continuum and line emission.
(a)
Map of the line core.
Black contours show the intensity distribution
averaged over the velocity interval of $V_{\rm LSR}$ = (5.8, 7.6) km s$^{-1}$.
Contour levels are 3, 4, and 5 times the rms noise (0.7 mJy beam$^{-1}$).
(b)
Maps of the line wings.
Blue and red contours are for the velocity intervals
of (--1.0, 5.8) and (7.6, 14.4) km s$^{-1}$, respectively.
Contour levels are 3, 4, 5, 6, 7, and 8 times the rms noise
(0.4 mJy beam$^{-1}$).
Some outflow peaks are labeled (also see Figure 5).
Gray scale:
map of the SiO $v$ = 0 $J$ = 1 $\rightarrow$ 0 line
(Figure 3a of Choi 2005).
Plus signs:
the 3.6 cm continuum sources (Reipurth et al. 2002).}
\end{figure*}

\begin{figure*}[!t]
\begin{center}
\FigureFile(162mm,0){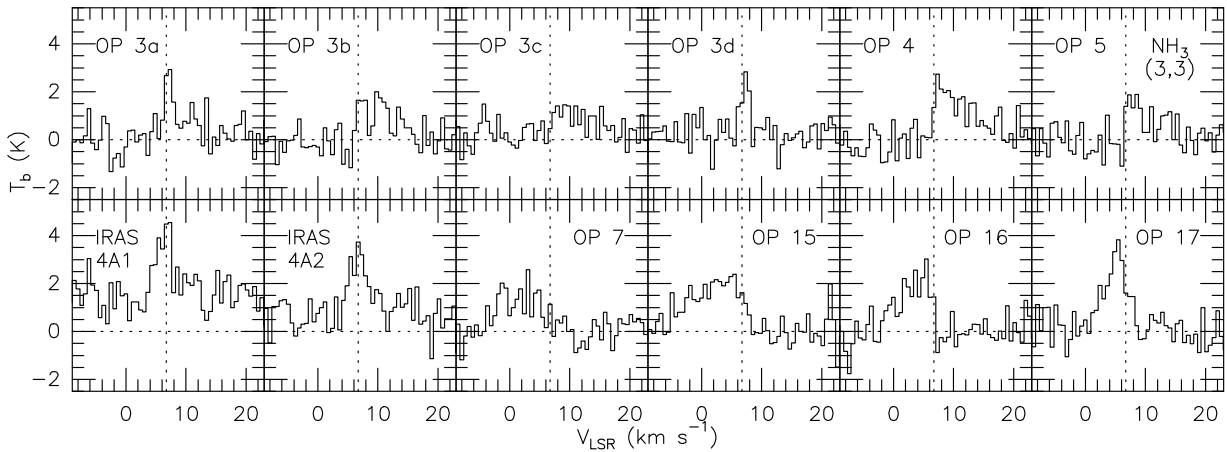}
\end{center}
\caption{
Spectra of the NH$_3$ (3, 3) line
from the C-array maps with an angular resolution of 2\farcs0 (Figure 6)
toward several NH$_3$ outflow peaks and IRAS 4A1/2.
Vertical dotted line:
systemic velocity of IRAS 4A.}
\end{figure*}

If A1 is indeed younger than A2, how large is the age gap?
To answer this question, the collapse age of A1 needs to be measured
using the rotation kinematics of the disk, as what was done with A2,
which requires imaging with a resolution higher than ours.
Instead, here we simply calculate a characteristic time scale of the system
as a sketchy expectation based on simple physics.
The age gap may be in the order of a time scale
that governs the dynamics of the A1-A2 system, i.e., a sound-crossing time.
Assuming an effective sound speed of 0.27 km s$^{-1}$ (Young \& Evans 2005),
the projected separation of 1\farcs8 (430 AU)
gives a sound-crossing time of $\sim$7,500 yr.
Considering that the A1-A2 system in not necessarily in the plane of the sky,
the value above is a lower limit.
For example, if we assume
that the A1-A2 system is $\sim$30$^\circ$ out of the plane of the sky,
the sound-crossing time would be $\sim$9,000 yr,
which is $\sim$20\% of the collapse age of A2.
(See Figure 5 for the comparison
between the binary separation and the infall region.)

\subsection{Outflows}

The molecular outflows in the IRAS 4A region
were detected in the NH$_3$ (3, 3) line
but almost undetected in the (2, 2) line.
Since some weak emission structures of the outflows
tend to be buried by noise
when the nominal synthesized beams were used,
the maps were convolved with larger beams to show the outflows clearly.
Figure 5 shows a schematic diagram of the IRAS 4A system,
and Figure 6 shows NH$_3$ (3, 3) line maps
with an angular resolution of 2$''$.
Figure 7 shows the spectra at outflow peak positions.
Figure 8 shows the NH$_3$ (3, 3) line map
with an angular resolution of 1$''$
for the velocity range showing the outflow peak (OP) 3a clearly.
Figure 8 also shows the SiO line map
showing the northeastern jet of IRAS 4A2.
For these maps (Figures 6 and 8),
contributions from the continuum emission were not removed
because the velocity extents of the line wings at strong peaks
are comparable to the spectral coverage
and left only a little room for line-free channels
that can be used for determining the continuum levels.
The structures directly associated with the continuum sources in these maps
should be ignored.

\begin{figure}[!t]
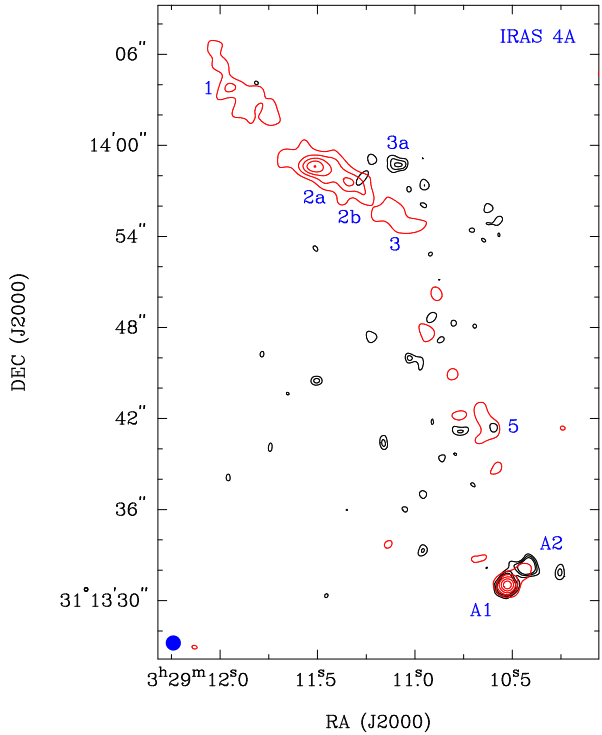

\begin{center}
\FigureFile(79mm,0){f08.eps}
\end{center}
\caption{
Maps of the northeastern area of the IRAS 4A region.
The continuum levels were not subtracted from the line data.
Black contours:
map of the NH$_3$ (3, 3) line from the VLA C-array data,
averaged over the velocity interval of $V_{\rm LSR}$ = (6.4, 7.6) km s$^{-1}$.
Contour levels are 3, 4, and 5 times the rms noise (0.5 mJy beam$^{-1}$).
Negative levels are not shown, to avoid clutter.
Shown in the bottom left-hand corner is the synthesized beam:
FWHM = 0\farcs98 $\times$ 0\farcs95 with PA = 47$^\circ$.
Red contours:
map of the SiO $v$ = 0 $J$ = 1 $\rightarrow$ 0 line from the VLA C-array data,
averaged over the velocity interval of (7.7, 25.9) km s$^{-1}$.
Contour levels are 1, 2, 3, 4, and 5 $\times$ 0.5 mJy beam$^{-1}$,
and the rms noise is 0.18 mJy beam$^{-1}$.
The SiO map was convolved to have an angular resolution of FWHM = 1\farcs0.
Protostars and outflow peaks are labeled.}
\end{figure}

In general, most of the NH$_3$ peaks associated with the IRAS 4A1/2 outflows
are located on or near the SiO outflow peaks (Figure 6b),
but the intensity distributions are quite different.
In the SiO outflow maps,
the strongest redshifted emission comes from OP 1 and 2,
and the strongest blueshifted emission comes from OP 7 and 9.
By contrast, in the NH$_3$ maps,
OP 3 and 4 are the strongest redshifted peaks,
and OP 15 and 16 are the strongest blueshifted peaks.

OP 17 is unusually strong in the NH$_3$ line,
but the nature of this source is not clear (Lefloch et al. 1998; Choi 2005).
In addition to the IRAS 4A outflows,
the NH$_3$ map also shows the outflow associated with IRAS 4BI
(the redshifted emission at $\sim$4$''$ north of BI in Figure 6b).

\subsection{Deflection of the Northeastern Outflow}

One of the most interesting properties of the IRAS 4A2 outflow is the directional variability of the northeastern (redshifted) jet.
The change of flow direction, or a sharp bend, near OP 3
is a good example of jet deflection
caused by a collision with a dense core in the ambient cloud (Choi 2005).
Figure 8 shows the detailed structure of the SiO jet,
and Figure 9 shows the spectra.
The undeflected upstream part (OP 3--5) is weak in the SiO line.
The deflected downstream part (OP 1--3) is relatively strong in the SiO line.
OP 2 in the low-resolution map (Choi 2005)
is now resolved into two peaks (OP 2a and 2b) in the new map.
Though these two peaks are spatially close to each other
(within 2\farcs5 or 600 AU),
their velocity structures are very different:
OP 2a has double peaks (one at 17 and the other at 21 km s$^{-1}$),
and OP 2b has a single peak (at 18 km s$^{-1}$).
This sudden change of velocity structure
suggests that the deflected jet is highly turbulent.

\begin{figure}[!b]
\begin{center}
\FigureFile(79mm,0){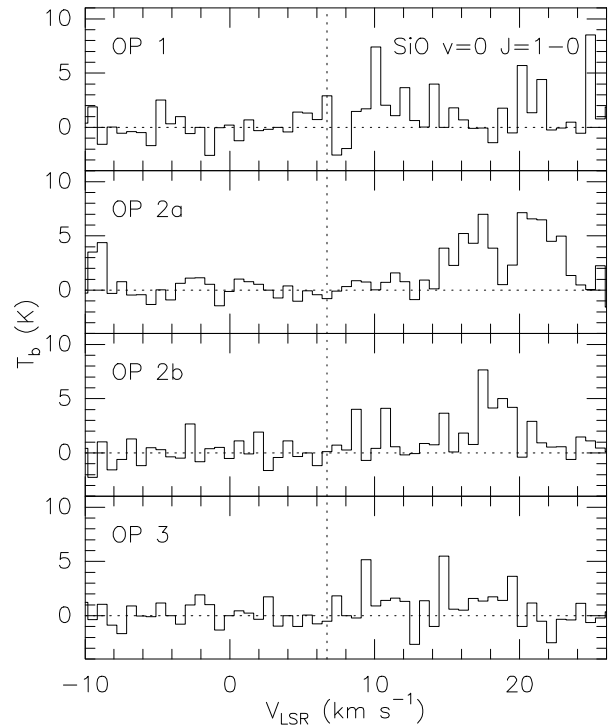}
\end{center}
\caption{
Spectra of the SiO line toward SiO outflow peaks 1--3
(see Figure 8 for the positions of the outflow peaks).
Vertical dotted line:
systemic velocity of IRAS 4A.}
\end{figure}

\begin{figure*}[!t]
\begin{center}
\FigureFile(78mm,0){f10a.nogray.eps}\hspace{5mm}
\FigureFile(78mm,0){f10b.nogray.eps}
\end{center}
\centerline{\scriptsize
[See http://minho.kasi.re.kr/Publications.html for the original figure.]}
\vspace{-\baselineskip}
\caption{
Maps of the molecular outflows in the IRAS 4A region.
For these maps, continuum levels were subtracted out.
(a)
Maps of the H$_2$CO 141 GHz line (Choi et al. 2004).
Blue and red contours show the intensity distribution
averaged over the velocity intervals
of $V_{\rm LSR}$ = (--1.0, 5.8) and (7.7, 14.4) km s$^{-1}$, respectively.
The lowest contour level and the contour interval are 70 mJy beam$^{-1}$,
and the rms noise is 22 mJy beam$^{-1}$.
Negative levels are not shown, to avoid clutter.
(b)
Maps of the HCN $J$ = 1 $\rightarrow$ 0 line (Choi 2001).
See Figures 1 and 2 of Choi (2001) for the velocity intervals.
Note that the blueshifted and redshifted line wings cannot be clearly separated
because of the blending of hyperfine components.
The lowest contour level and the contour interval are 60 mJy beam$^{-1}$,
and the rms noise is 20 mJy beam$^{-1}$.
Gray scale:
map of the SiO line (Figure 3a of Choi 2005).
Plus signs:
the 3.6 cm continuum sources (Reipurth et al. 2002).}
\end{figure*}

One of the motivations of the new high-resolution observations
was imaging the detailed structure of the jet at the jet-core impact point
that is supposed to be near OP 3.
However, the SiO emission of OP 3
is weak and does not show a compact structure,
which may mean that the SiO emission
does not get bright immediately after the impact
but takes some time to develop.
A likely explanation is
that Si atoms in dust grains
are injected into the shocked gas at the impact point,
combine with oxygen in the gas,
and form SiO in a time scale of 100--1,000 yr
(Schilke et al. 1997; Gusdorf et al. 2008; Guillet et al. 2009).
The SiO molecules then combine again with oxygen and become SiO$_2$.
Therefore, the strength of the SiO emission would peak
with a certain time lag after the jet-core impact.

In the deflected jet of IRAS 4A2,
the SiO emission is brightest at OP 2a, $\sim$7$''$ from the sharp bend.
There is no measurement of the proper motion in this part of the outflow.
If we use the proper motion measured in the southwestern jet
(0\farcs064 yr$^{-1}$; Choi et al. 2006) as a very rough guide, 
the time scale of outflow propagation from the impact point to OP 2a
is $\sim$100 yr.
The argument above assumes that the jet is a steady flow,
but images of the IRAS 4A2 jet show
that the brightness distribution is far from uniform.
The brightness distribution of the deflected jet
would also be subject to the non-steady structure of the impacting jet.
Considering the average separation between outflow peaks,
the uncertainty in the time scale of SiO peak emission would be $\sim$40\%.

The NH$_3$ maps (Figures 6 and 8) show an interesting structure around OP 3. 
While the emission peaks directly associated with the outflow (OP 3b and 3c)
show wide redshifted line wings,
those located slightly away from the SiO jet (OP 3a and 3d)
show narrow lines near the ambient velocity (Figure 7).
OP 3a and 3d are probably parts of the dense core
that is obstructing the northeastern jet and causing the deflection.
NH$_3$ molecules at OP 3a is excited
either by the shock propagated from the impact point
or by a stream of jet that penetrated deep into the obstructing core.
Examinations of the centimeter-millimeter continuum maps show
that there is no detectable continuum source at or near the position of OP 3a
(Rodr{\'\i}guez et al. 1999; Reipurth et al. 2002;
Choi et al. 2004, 2007; this work).

\subsection{Outflow Chemistry}

As mentioned above, there is an interesting difference
between the intensity distributions of SiO and NH$_3$.
Comparisons with other outflow tracers would be helpful
in understanding the intensity distributions.
Figure 10 shows the H$_2$CO and HCN maps of the IRAS 4A outflows
(Choi 2001; Choi et al. 2004).
The outflow structure in SiO is most well-collimated and straight,
and the outflow images in the other tracers agree with the SiO image
in the overall structure.
Outflow peaks in the four tracers
are usually located close to each other.
Especially, OP 1, 3, 4, and 15 are detectable in all the four outflow maps,
and the peak positions of each OP in these maps
are clustered within $\sim$2$''$.
(The average separation between adjacent peaks is $\sim$6$''$.)
This clustering suggests
that a single physical agent (e.g., internal shock in the primary jet)
is responsible for the excitation of all the four outflow tracers. 
Therefore, the outflowing gas components traced by these tracers
are physically closely connected to each other
and flow with a certain degree of coherence.

\begin{figure*}[!t]
\begin{center}
\FigureFile(162mm,0){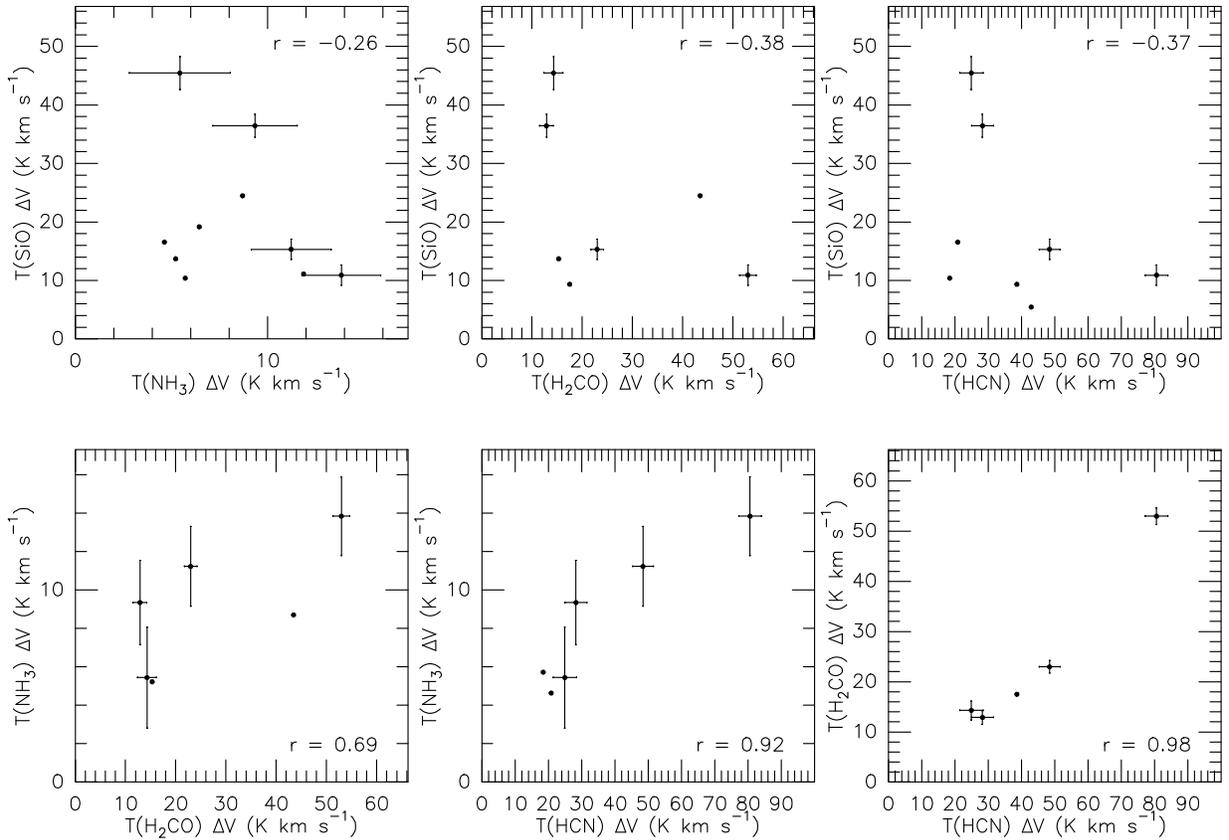}
\end{center}
\caption{
Correlation diagrams of the SiO, NH$_3$, H$_2$CO, and HCN line intensities
for outflow peaks (see the maps in Figures 6b, 10a, and 10b).
The integrated intensities are corrected for the primary beam response.
Uncertainties are marked for outflow peaks 1, 3, 4, and 15.
Written in the top/bottom right-hand corners
are the linear correlation coefficients.}
\end{figure*}

Though the outflow tracers share the source of excitation,
SiO stands out when the intensity distributions are compared.
In the northern part of the IRAS 4A region,
the SiO map shows
that the deflected (downstream) part of the northeastern jet (OP 1--3)
is brighter than the undeflected (upstream) part (OP 3--5).
The maps of the other tracers show that the upstream part is brighter.
In the southern part of the region,
the SiO map shows
that the southwestern jet (blueshifted jet of IRAS 4A2: OP 7--12)
is brighter than the southern outflow
(blueshifted outflow of IRAS 4A1: OP 13--16).
The maps of the other tracers
show that the southern outflow (especially OP 15 and 16) is brighter.
The intensity distribution pattern of the NH$_3$, H$_2$CO, and HCN maps
is similar to that of the CO $J$ = 2 $\rightarrow$ 1 outflow map
(see Figure 1 of Girart et al. 1999).

To show this trend quantitatively,
correlation diagrams of the outflow peaks were made.
The intensity of the SiO line tends to be anti-correlated
with those of the other lines (Figure 11, top panels).
The linear correlation coefficients are negative: $r \approx -0.3$.
If only the outflow peaks detected in all the four lines are considered
(marked with uncertainty bars in Figure 11),
the anti-correlation is even stronger: $r\approx -0.9$.
By contrast, the intensities are positively correlated
among NH$_3$, H$_2$CO, and HCN: $r \approx 0.9$ (Figure 11, bottom panels).
This difference suggests
that the outflow component traced by SiO
is fundamentally different from that of the other tracers.

Blake et al. (1995) arrived at a similar conclusion
based on the spectral profiles of molecular lines
obtained from single-dish observations,
which show that the SiO emission
is kinematically displaced from the bulk cloud velocity.
The image and spectra of the SiO line suggests
that the gas traced by the SiO emission is
either the primary jet itself
or an outflow component closely linked to the primary jet,
as suggested by the observations of other outflows such as L1448 and HH 212
(Dutrey et al. 1997; Codella et al. 2007).

The dichotomy between SiO and the other outflow tracers
can be explained in two different ways.
First, the differences in shock strength
may trigger different chemical processes.
While the sputtering of Si from dust grains
requires shock velocities higher than $\sim$25 km s$^{-1}$,
species such as H$_2$CO do not survive at such high velocities
(Blake et al. 1995; Schilke et al. 1997; Arce et al. 2007).
In this scenario, the intensity ratio of SiO to the other species
can be an indicator of the shock strength.
Second, the line intensities may be tracing
either different physical properties
or different regions of the outflow system.
While SiO may reflect the shock strength of the jet itself,
the other tracers may reflect physical conditions of the ambient medium
such as the density of ambient cloud (Blake et al. 1995).

\section{SUMMARY}

The NGC 1333 IRAS 4A region was observed
using the VLA in the 1.3 cm and 6.9 mm continuum,
the NH$_3$ (2, 2) and (3, 3) lines,
and the SiO $v$ = 0 $J$ = 1 $\rightarrow$ 0 line,
with an angular resolution of $\sim$0\farcs4,
to image the circumstellar structures and outflows
of the protobinary system.
The continuum emission mostly comes
from dust in the circumstellar disks and the protostellar envelopes.
The high-resolution (0\farcs3) images of the NH$_3$ lines
show the detailed structures of the disks and outflows around the protostars.
The molecular line images of relatively low resolutions (1$''$--2$''$)
were used to investigate the properties of the protostellar outflows.
The main results are summarized as follows.

1.
The continuum maps show two compact emission sources
and a clumpy extended structure surrounding them.
They may include the circumstellar disks, protostellar envelopes,
and the common envelope of the binary system.
The steep SED suggests
that most of the 1.3 cm and 6.9 mm continuum emission comes from dust.
The mass of the dense molecular gas in the binary system,
imaged by the millimeter-continuum interferometric observations,
is $\sim$2.6 $M_\odot$.
The IRAS 4A1 protostar-disk-envelope system
is more massive than the A2 system by a factor of $\sim$6.

2.
The NH$_3$ line maps of the region immediately around IRAS 4A1
show at least two components.
One is probably the circumstellar disk,
and the other may be the blueshifted southern outflow.
The blueshifted and redshifted emission sources
of the central (disk) component
was not spatially separated,
suggesting that the disk may be rotating relatively slowly.
Analysis of the rotation curve, as what was done with the A2 disk,
was not possible with the angular resolution of this work,
and suggests that the A1 protostar is less massive than the A2 protostar.

3.
Considering the properties of the disks and the lengths of the outflows,
IRAS 4A1 may be younger than A2, in terms of collapse age.
We suggests that the less massive member of a binary prestellar core system
may start the protostellar collapse earlier than its companion.

4.
The molecular outflows are brighter in the NH$_3$ (3, 3) line
than in the (2, 2) line.
The overall structure of the NH$_3$ outflows
is similar to that of the SiO jets/outflows,
though there are interesting differences
in small-scale structures and intensity distributions.

5.
The 1$''$ resolution map of the SiO line shows
the highly turbulent structures
of the deflected part of the IRAS 4A2 northeastern jet.
The SiO line is brightest at a peak located $\sim$7$''$
from the starting point of the deflected part,
suggesting that the enhancement of SiO in the jet
takes $\sim$100 yr after the proposed jet-core collision.
The NH$_3$ line maps
show an ambient-velocity structure near the impact point,
which may be a part of the obstructing dense cloud core.

6.
Comparisons of the SiO, NH$_3$, H$_2$CO, and HCN maps of the IRAS 4A outflows
show an interesting chemical dichotomy.
The intensity distribution of SiO
is anti-correlated with those of the others.
This trend suggests
that NH$_3$, H$_2$CO, and HCN may trace
the same bulk of gas in the molecular outflows,
while SiO traces a gas component more directly related to the primary jet.

\bigskip

\enlargethispage{-5\baselineskip}

We thank Jongsoo Kim for helpful discussions.
M. C. and M. K. were supported by the Core Research Program
of the National Research Foundation of Korea (NRF)
funded by the Ministry of Education, Science and Technology (MEST)
of the Korean government (grant number 2011-0015816).
J.-E. L. was supported by the Basic Science Research Program
through NRF funded by MEST (grant number 2011-0004781).
The National Radio Astronomy Observatory is
a facility of the National Science Foundation
operated under cooperative agreement by Associated Universities, Inc.

\vfill
\centerline{\small\tt arXiv version}
\vspace*{-\baselineskip}

\end{document}